\documentclass[12pt,a4paper]{article}

\usepackage[utf8]{inputenc}
\usepackage[T1]{fontenc}
\usepackage{lmodern}
\usepackage[english]{babel}
\usepackage{amsmath,amsfonts,amssymb}
\usepackage{graphicx}
\usepackage[margin=1in]{geometry}
\usepackage{setspace}
\usepackage{natbib}
\usepackage{url}
\usepackage{hyperref}
\usepackage{xcolor}
\usepackage{soul}

\usepackage{multirow} 
\usepackage{array}

\usepackage{textgreek}
\DeclareUnicodeCharacter{03BC}{\textmu}

\hypersetup{
    colorlinks=true,
    linkcolor=blue,
    filecolor=magenta,      
    urlcolor=cyan,
    citecolor=blue,
}

\title{\textbf{Electroencephalography and Electromyography as a Non-Invasive Biomarker of Neural Regeneration: A Review of Central and Peripheral Nervous System Injury and Regeneration}}

\author{
Maryam Kheyrollah\textsuperscript{1}, Reza Khanbabaie\textsuperscript{1, 2}, Chris Ullrich \textsuperscript{3},\\ Mohammad Moulaeifard\textsuperscript{4,*}\\[0.9em]
\textsuperscript{1}Department of Physics, University of Ottawa, Ottawa, Canada\\
\textsuperscript{2}Congnixion Inc., Toronto, Canada\\
\textsuperscript{3}Congnixion Inc., Santa Barbara, CA, USA\\
\textsuperscript{4}Department of AI4 Health, University of Oldenburg, Oldenburg, Germany\\[0.5em]
\textsuperscript{*}Corresponding Author
}

\date{}

\begin{document}

\maketitle

\section*{Abstract}

Regeneration of the nervous system after injury remains an important therapeutic objective, especially in the central nervous system (CNS), in which regeneration is restricted by both neuronal limitations as well as adverse extracellular environments. Conversely, the peripheral nervous system (PNS) displays enhanced regenerative capability in the presence of supportive Schwann cells (SC) and pro-growth stimuli. While the structure and molecular mechanisms are thoroughly understood, functional biomarkers that can non-invasively monitor regeneration in real time are limited. In this review, we discuss the promise of electroencephalography (EEG) as well as electromyography (EMG) as real-time, non-invasive biomarkers to monitor damage to nerves and regeneration in both CNS and PNS contexts.

First, we contrast biological and electrophysiological indicators of CNS/PNS injury, showing how EEG signs, including oscillatory power, connectivity, and evoked potential changes, reflect dysfunction due to injury as well as neuroplastic reorganization. Also, EMG provides direct insight into muscle activation and peripheral output, providing useful EEG complementation in neuromuscular pathway integrity and reactivation. In CNS injuries (e.g., stroke, spinal cord injury (SCI)), EEG typically shows global slowing, disrupted interhemispheric coherence, and partial recovery of higher frequencies. For PNS injuries, EEG can capture cortical remapping and return of somatosensory evoked responses with re-establishment of the peripheries' connectivity. EMG, in turn, enables monitoring of reinnervation and restoration of functional motor output.

This review presents a dual-system perspective, positioning EEG and EMG not only as diagnostic tools but also as functional biomarkers of neural regeneration, thereby bridging electrophysiology, plasticity, and clinical recovery.

\section{Introduction}

Injuries of the nervous system elicit vastly different regenerative responses within central and peripheral compartments. The adult mammalian CNS has a severely limited capacity for repair, often resulting in persistent neurological impairments. In contrast, the PNS has a significant capacity for axonal regrowth, which can yield significant functional recovery \citep{cooke2022neuronal}. Given these disparities, strong interest has developed in the development of credible, non-invasive biomarkers of neural regeneration and their use in guiding processes of rehabilitation. EEG has emerged as a promising tool in this regard, and EMG is now seen as a supplementary tool for recording peripheral reinnervation and motor recovery. EEG and EMG are both non-invasive measures that register electrical activity in the brain and skeletal muscles, respectively, therefore granting immediate access to real-time information about neural functioning. Moreover, these approaches are fairly available and inexpensive, features which have helped fuel their use as electrophysiological measures of neurological recovery and predictive measures of functional outcome \citep{shigapova2024electrophysiology}.

EEG itself measures the brain's electrical rhythms directly and can mirror neuroplastic adaptations underlying recovery following injury. Quantitative EEG measures have been related to neuroplasticity within diverse CNS injuries. For instance, in patients with SCI, EEG spectral power changes (e.g., the alpha/theta ratio) have been related to locomotor and balance function improvement during a course of rehabilitation \citep{simis2021electroencephalography}. Similarly, EEG measurements of cortical activation and connectivity obtained after stroke have also been related to motor outcomes \citep{delcamp2024eeg}. EEG has several practical advantages within these scenarios; it is nonharmful, bedside status ready, and is capable of documenting dynamic brain network reorganization after extended time periods. By identifying cortical markers of adaptive plasticity, EEG is an important marker of central recovery \citep{bareham2018longitudinal}.

EMG, by contrast, measures muscle electrical activity and offers a window into the integrity of the spinal and peripheral nerves \citep{sun2020complexity}. For peripheral nerve injury, EMG is a pillar used to assess the severity of nerve injury and follow regeneration \citep{orthobullets_peripheral}. When regenerating axons reinnervate muscle fibers, EMG can pick up on the recovery of motor unit potentials (MUPs) and, essentially, document reinnervation by collateral sprouting and axonal growth \citep{lam2024peripheral}. For CNS injuries with motor impairment, EMG is used advantageously to follow the return of voluntary muscle function \citep{sharma2023preservation}. EMG has been utilized within models of SCI to measure restored locomotor function at recovery \citep{pilkar2020use}. Accordingly, although EEG measures central neural plasticity, EMG measures peripheral muscle activation at the effector level and hence the end product of neural regeneration \citep{balbinot2022surface}.

Notably, some neurophysiological tests combine EEG and EMG to evaluate the connectivity of pathways and functional transmission. A classic illustration is evoked potential testing, where somatosensory evoked potentials (SSEPs) are recorded from the cortex using EEG, and motor evoked potentials (MEPs) are recorded using EMG from target muscles. These tests objectively measure the integrity of ascending and descending pathways, respectively. Notably, improvements in SSEP and MEP responses have been shown to parallel functional motor recovery after neural injury. For instance, after an incomplete spinal cord lesion, progressive enhancement of SSEP and MEP amplitudes (and decrements in latencies) with time are associated with better locomotor and functional outcomes. Such observations show how EEG/EMG-based integrated measures can monitor the recovery of neural circuits along the injury continuum, from cortical signal origination to muscle activation \citep{li2021utility}.

\section{Nerve Injury and Regeneration in the CNS and PNS}

Nerve injuries to the PNS and CNS vary considerably in their biological potential for repair and in how we follow functional recovery. Most often, CNS neurons do not regenerate after being injured because of an inhibitory environment, for instance, glial scars and myelin-associated inhibitors actually retard axon growth inside the brain and spinal cord \citep{silver2004regeneration, varadarajan2022central}. By contrast, PNS possesses a more favorable milieu for regeneration. Following peripheral nerve injury, SCs remove myelin debris and release growth factors, directing axonal sprouts across the injury distance \citep{lutz2017schwann, deumens2010repairing}. These fundamental differences suggest that CNS injuries often result in lasting deficits, whereas PNS injuries can, at times, achieve remarkable reinnervation and functional restitution under proper conditions. Beyond the biological mechanisms, there are also differences in the neurophysiological biomarkers of injury and recovery in each system \citep{pleasure1999regeneration, kuffler2020restoration}. EEG has been comprehensively applied to explore CNS injuries and brain activity changes \citep{rayi2020electroencephalogram, feng2025electroencephalography}, while EMG represents a companion method to examine muscle activation and, therefore, neural circuitry output in both PNS and CNS injuries \citep{brambilla2021combined}. In what follows, we contrast PNS and CNS injuries, underscoring both distinctions between their regenerative response and also EEG/EMG measurements for tracing neurological and functional recovery.

\subsection{CNS Injury and Regeneration}

In the adult CNS, traumatic injury commonly triggers a cascade of events that inhibits axonal regeneration. Damaged CNS axons exhibit an intrinsic lack of robust regenerative capacity, and this is compounded by extrinsic factors in the CNS environment \citep{huebner2009axon, shafqat2023tackling}. In addition, myelin-associated inhibitory molecules like Nogo, MAG, and OMgp that are released from oligodendrocytes and myelin debris activate repulsive signaling mechanisms that result in growth cone collapse and inhibit new extension \citep{chambel2023axonal, mckerracher2015mag}. Taken together, these intrinsic and extrinsic barriers create a highly inhibitory environment that is responsible for the restricted regenerative capacity of the CNS.

As a result, neurons above and below the lesion are often unable to re-establish connections and thus exhibit chronic neurological deficits \citep{fink2016reorganization, kvistad2024neural}. By contrast, peripheral nerves that have been severed experience Wallerian degeneration and cleaning up of debris, followed by growth-supportive SC forming Bungner bands to direct growing axons \citep{stoll1999nerve, panzer2020tissue}. There is no similar supportive process in the CNS, in fact, CNS myelin and scar tissue actually retard regeneration \citep{gaudet2011wallerian, fu2025research}. Thus, from a biological standpoint, CNS injuries are generally permanent, whereas PNS injuries have at least a chance of anatomical and functional recovery \citep{balakrishnan2021insights}.

Despite the limited regenerative capacity of the adult CNS, significant neuroplasticity can occur after the injury. Surviving circuits often restructure, and adjacent cortical areas can assume the functions of damaged regions. Meanwhile, spared pathways, such as the corticospinal, reticulospinal, or propriospinal tracts, can be reinforced to compensate for lost connections \citep{filipp2019differences, gao2022restoring}. EEG enables non-invasive monitoring of cortical reorganization, as indicated by altered oscillation patterns (such as slowed alpha rhythms and altered beta-band activity), disrupted functional connectivity, and atypical event-related potentials during motor tasks in stroke or SCI patients \citep{lacerda2024eeg, li2020eeg}. EEG was used to monitor cortico-cortical connectivity and cortico-muscular coherence during motor rehabilitation, finding that specific EEG metrics change as patients recover movement \citep{krauth2019cortico}. Such EEG biomarkers present a non-invasive perspective on the reorganization of the brain as well as the integrity of the CNS pathways.

As a complement to EEG, EMG provides crucial insight into the motor output capacity after a CNS injury. In upper motor neuron lesions (above the segmental level), the peripheral nerves and muscles are not directly injured; hence, EMG at rest typically shows no denervation potentials, unlike in PNS injuries \citep{benecke1983denervation}. Instead, the hallmark of a pure CNS lesion on EMG is impaired voluntary recruitment of motor units. In a complete SCI (with no descending signals), voluntary MUPs are missing on EMG recordings below the lesion. In contrast, in incomplete injuries, however, certain MUPs remain evocable, frequently at lower numbers but with normal configuration and amplitude, indicating partly intact pathways \citep{robinson2022traumatic}. Therefore, the presence of any volitional EMG activity in muscles below a CNS lesion is an important indicator of residual descending motor tract connectivity. Indeed, sensitive EMG recordings have revealed "hidden" residual voluntary output in some patients diagnosed as clinically complete SCI. In one study, surface EMG revealed subtle but consistent muscle activations in a majority of chronic motor-complete SCI subjects when they attempted movements, indicating that below-injury volitional muscle activity was present in many cases \citep{heald2017characterization}. These findings demonstrate how EMG can reveal spared motor units and descending signals not seen in clinical exams.

EMG is a valuable tool for determining the integrity of the descending pathways. Via transcranial magnetic stimulation, MEPs may be evoked and verified on EMG, providing evidence of corticospinal conduction even through spinal lesions. In those with complete paralysis, MEPs mainly appear in individuals who develop spasticity, suggesting preserved descending pathways linked to spastic activity \citep{sangari2019residual}. This implies that when EMG records muscle activity (voluntary or evoked) below a CNS lesion, it indicates sparing of motor axons and serves as a positive prognostic marker, as voluntary EMG is highly sensitive in detecting preserved motor function post‑SCI \citep{santamaria2021neurophysiological}. Moreover, EMG can characterize abnormal activation patterns, such as involuntary spasms and abnormal co‑contractions that frequently accompany spasticity \citep{balbinot2022surface, sheng2022upper}. While EEG reflects brain function, EMG monitors muscle activation from neural signals, complementing each other in CNS injuries by showing cortical changes and motor execution status, thus revealing the integrity of descending pathways \citep{debenham2024neuromuscular}.

\subsection{PNS Injury and Regeneration}

Peripheral nerve injuries (such as brachial plexus traction, nerve lacerations, or compression neuropathies) have a very different healing response and recovery profile compared to CNS injuries. The PNS axons regenerate at the rate of 1-3mm/day when distal nerve circuits (endoneurial tubes) are maintained \citep{carvalho2019modern, sulaiman2013neurobiology}. After peripheral nerve damage, the distal segment experiences Wallerian degeneration \citep{rotshenker2011wallerian}. In this process, the axon disintegrates, and the resulting myelin is removed by phagocytes. SC dedifferentiate, proliferate, line up inside endoneurial tubes (to form Bands of Büngner) \citep{nevmerzhytska2025review}, and produce growth factor molecules (e.g., NGF, BDNF) to guide growing axons \citep{rotshenker2011wallerian, nevmerzhytska2025review}. If the gap between nerve stumps is not too large (or is surgically bridged), axonal sprouts from the proximal stump can enter the distal Schwann cell channels and eventually reinnervate target muscles and sensory receptors \citep{sulaiman2013neurobiology}. Functional recovery is not complete, however, due to misdirection of regenerating axons \citep{nguyen2002pre}, degeneration of the motor endplates \citep{huang2022denervation}, and denervation atrophy \citep{carlson2014biology} Despite these circumstances, the inherent regenerative capability of PNS neurons and favorable extracellular environment mean that, unlike in the CNS, there is genuine potential for useful partial recovery following numerous peripheral nerve injuries \citep{grinsell2014peripheral, gordon2020peripheral}.

Electrophysiologic methods play an important role in the development of peripheral nerve regeneration. While PNS injuries are outside of the brain, EEG-based methods are less important in these cases; however, there are instances where EEG can play a role. For example, SEPs can be utilized by the clinician to evaluate the recovery of the sensory pathway after repair. By electrically stimulating the injured nerve and recording EEG responses over the somatosensory cortex, the reappearance of cortical evoked potentials indicates the successful regeneration of sensory fibers across the lesion \citep{synek1987role}. Cortical SEPs appear before peripheral nerve responses due to central amplification \citep{eisen1988use}. Accordingly, EEG SEP testing can confirm early access of afferent signals to the brain, indicating physiological reconnection in the PNS before peripheral outcomes become visible \citep{synek1987role}. EEG indicates cortical reorganization following chronic peripheral injury via motor cortex mapping changes or different motor imagery signals, and central plasticity is of secondary significance \citep{amini2018peripheral, li2021cortical}. EMG patterns of reinnervation and recovery of the muscle, such as the progression from small nascent motor units to stable polyphasic MUPs, reflect the peripheral restoration of the nerve directly \citep{krarup2016remodeling}.

Early after a severe PNS injury (axonotmesis or neurotmesis), there is a characteristic sequence of EMG changes. In the first 1-2 weeks following injury, the motor nerve and motor fibers may still conduct a few impulses due to the remaining acetylcholine at the motor endplate and the delay of Wallerian degeneration, so EMG and nerve conduction studies may show little or no change initially \citep{panzer2020tissue, chaney2020axonotmesis}. 2-3 weeks following injury, after Wallerian degeneration is completed and the muscles have been denervated, the EMG typically reveals strong patterns of acute denervation, and the respective muscles demonstrate spontaneous fibrillation potentials and positive sharp waves on rest, indicating that the muscle fibrils have lost innervation. These spontaneous potentials usually appear once axonal degeneration reaches the motor end plate and are well-recognized markers of denervation during the electrodiagnostics \citep{ginsberg2020using, stalberg2019standards}. There are initially no MUPs present within completely denervated muscles, whereas in partial injuries, recruitment is markedly reduced. This EMG profile provides early confirmation of axonal loss and is central to distinguishing neuropraxia where no fibrillations can be distinguished and rapid recovery is achieved, from axonotmesis or neurotmesis, where regrowth of the axons is required \citep{kamble2019peripheral, preston2012electromyography}.

\begin{table}[htbp]
\centering
\caption{EEG and EMG Biomarkers in CNS and PNS Injury and Recovery}
\label{tab:eeg_emg_biomarkers}
\resizebox{\textwidth}{!}{%
\begin{tabular}{|p{2.5cm}|p{3.5cm}|p{3.5cm}|p{3.5cm}|p{3cm}|}
\hline
\textbf{System} & \textbf{EEG Findings} & \textbf{EMG Findings} & \textbf{Recovery Markers} & \textbf{Clinical Applications} \\
\hline
\multirow{5}{2.5cm}{\textbf{CNS Injury} \\(Stroke, TBI, SCI)} 
& Global slowing (increased theta/delta) & Reduced voluntary motor unit recruitment & Spectral normalization (decreased delta/theta, increased alpha/beta) & Prognosis prediction \\
\cline{2-5}
& Disrupted interhemispheric coherence & Abnormal co-contractions & Enhanced movement-related ERD/ERS & Rehabilitation monitoring \\
\cline{2-5}
& Altered oscillatory patterns & Spastic muscle activity & Restored cortico-muscular coherence & BCI-FES systems \\
\cline{2-5}
& Reduced functional connectivity & Preserved MUPs in incomplete lesions & Improved interhemispheric connectivity & Neurofeedback therapy \\
\cline{2-5}
& Asymmetric hemispheric activity & Denervation potentials in severe cases & Recovery of voluntary EMG activity & Treatment adaptation \\
\hline
\multirow{5}{2.5cm}{\textbf{PNS Injury} \\(Peripheral nerve lesions)} 
& Minimal global changes (intact cortex) & Fibrillation potentials (2-3 weeks post-injury) & Reappearance of cortical SEPs & Early regeneration detection \\
\cline{2-5}
& Cortical remapping in chronic cases & Positive sharp waves & Polyphasic MUPs & Surgical timing decisions \\
\cline{2-5}
& Altered SEPs (delayed/absent) & Loss of voluntary MUPs & Progressive MUP maturation & Prosthetic control \\
\cline{2-5}
& Sensorimotor network reorganization & Complete denervation patterns & Increasing recruitment strength & Muscle retraining \\
\cline{2-5}
& Motor imagery changes & Collateral sprouting evidence & Stabilization of waveforms & Functional assessment \\
\hline
\multirow{3}{2.5cm}{\textbf{Combined Measures}} 
& \multicolumn{2}{p{7cm}|}{Cortico-muscular coherence (CMC) - measures brain-muscle coupling} & Increased beta-band CMC during recovery & Motor rehabilitation \\
\cline{2-5}
& \multicolumn{2}{p{7cm}|}{Movement-related cortical potentials (MRCPs) with concurrent EMG} & Synchronized EEG-EMG responses & Hybrid BCI development \\
\cline{2-5}
& \multicolumn{2}{p{7cm}|}{MEPs and SSEPs} & Improved latencies and amplitudes & Pathway integrity assessment \\
\hline
\end{tabular}%
}
\end{table}

These features reflect early reinnervation, and every polyphasic MUP is a motor unit that is reinnervated by a sprouting axon, but with immature neuromuscular junctions and remodeling nerve terminals, leading to variability in the waveform \citep{kamble2019peripheral}. Over time, as the maturing regenerating axons achieve more stable synaptic connections on muscle fibers, the MUPs change: widths expand, durations become longer, and waveforms become less polyphasic (more stable) as the motor units regain a more normal state. For instance, in the first few months, one may capture increasingly larger-amplitude and longer-duration MUPs in a recovering muscle and reflect the fact that the regenerating axons have reinnervated more fibers and neuromuscular transmission effectiveness has increased. This continuum on EMG, from lost voluntary MUPs to the onset of small polyphasic nascent units and ultimately to larger more stable motor units, offers an in vivo electrophysiological timetable of peripheral nerve regeneration. Serial EMG and nerve conduction studies offer pertinent prognostic information in PNS injuries clinically. The onset of reinnervation potentials on EMG offers early evidence of recovery often preceding noticeable muscle contraction by several weeks. For instance, EMG may demonstrate nascent polyphasic MUPs produced by collateral sprouting well before patients experience discernible voluntary strength \citep{stalberg2019standards, kamble2019peripheral, preston2012electromyography}. When no arising MUPs appear on EMG after several months, failed regeneration or too great a nerve gap exists and surgical repair such as nerve grafting may be necessary \citep{sulaiman2013neurobiology, kimura2025electrodiagnosis}. EMG also differentiates neuropathic and myopathic recovery by evaluating motor unit morphology and recruitment patterns \citep{campbell2005dejong}. Additionally, EMG can measure the quality of neuromuscular transmission: immature reinnervated motor units display jitter and variability at first, however over time interference patterns become stabilized as synapses mature \citep{stalberg2019standards}. Overall, EMG offers direct evidence of the recovery of motor units in peripheral injuries and persists as the gold standard for the observation of nerve regeneration in conjunction with the clinical examination \citep{kimura2025electrodiagnosis}.

Keep in mind that EEG primarily reveals what occurs in the brain, and the EMG reveals what occurs in the muscle. Together, they offer the whole picture of nerve injury recovery \citep{makeig2004mining}. This holds particularly for the PNS: the brain may desire to move (typical EEG activity for the desire to move), but only when peripheral nerve fibers begin growing again will the EMG in the muscle demonstrate genuine movement \citep{farina2014extraction}. The EEG and the EMG complement each other when monitoring recovery in the muscles, one identifies the brain's signals, and the other verifies the muscles' response. More research in neurorehabilitation incorporates both EEG and EMG recordings precisely for this reason. For instance, examination of EEG and EMG concurrently may quantify the degree to which the brain may control muscles following injury. The combined measure has been proposed as a good indicator of the recovery of motor function in individuals who've had strokes or spinal cord injuries \citep{wijk2012neural, mima1999corticomuscular, grosse2002eeg}. Table~\ref{tab:eeg_emg_biomarkers} summarizes the key findings of EEG and EMG biomarkers in CNS and PNS injury and recovery.

\section{EEG/EMG Changes Following CNS Injury}

Following CNS trauma, such as stroke, traumatic brain injury (TBI), or SCI, EEG tracings demonstrate unique disturbances. One common feature of such injuries is the slowing of the background rhythms. As a case in point, TBI patients will often present a deceleration of the posterior dominant rhythm and enhancement of diffuse theta activity shortly after the trauma. Similarly, in the case of stroke or encephalopathy, diffuse theta and delta slowing provide evidence for global cerebral dysfunction \citep{non_epileptiform2021slowing, ianof2017traumatic}. Following ischemic stroke, the electroencephalogram almost invariably demonstrates a decrease in the dominant alpha rhythm (8-12 Hz) and the appearance of slow delta and theta activity in the damaged areas, indicating acute dysfunction in the cortex \citep{peterson2025exploring}. These slow-wave defects reflect dysfunction in the cortex and impaired neural circuitry in the damaged brain \citep{fanciullacci2017delta}. Supporting the above findings, quantitative EEG measures routinely demonstrate that increased enhancement of slow waves and reduced power in the fast waves predict adverse clinical outcomes \citep{ajcevic2021early, sutcliffe2022surface}. Severe cases indeed demonstrate focal brain lesions and corresponding localized EEG suppressing or localized slowing by asymmetry of the hemispheres \citep{focal_abnormalities2025}.

Spectral slowing apart, the dynamics of brain networks are also impaired by CNS lesions. Focal lesion such as acute stroke induce global functional connectivity alterations outside the lesion site \citep{alia2017neuroplastic}. Acute stroke EEG functional networks less intensely connected become less integrated: the ipsilesional hemisphere shows lower alpha- and beta-band connectivity as compared to the contralesional hemisphere \citep{zhang2024resting}. SCI (lesion of a tract in the CNS) also induces cortical EEG changes despite the intactness of the brain tissue. Chronic SCI patients have exhibited decreased EEG alpha power and lower peak frequency and paradoxically increased beta power. Ascending sensory and descending motor output loss in SCI also commonly result in absent or transformed SSEPs on EEG \citep{doruk2017investigation}.

EMG provides valuable information regarding the peripheral manifestation resulting from the disruption of CNS lesions, including stroke, TBI, and SCI, on the motor pathways. In acute strokes, needle EMG characteristically shows evidence for lower motor neuron dysfunction in the paretic muscles even when the lesion is only in the brain. For example, about 40\% of the patients show denervation potentials, marked by the presence of positive sharp waves and fibrillations, along with abnormal recruitment patterns in the motor units on EMG at 72 hours after stroke. Such indices of denervation confirm that severe strokes do indeed induce secondary motor neuron or disuse atrophy and are associated with larger deficits and weakness at onset \citep{bitencourt2022needle}. In SCI, disruption of the descending control inputs by the brain may at times be unearthed by high-sensitivity surface EMG even when frank overt movement by the muscle is not evident, for instance, muscles graded as zero on the MMT. Such evidence advocates the utility using EMG to gauge the status of incompletely injured tissues \citep{balbinot2022surface, heald2017characterization}. Finally, chronic SCI causes typical alterations in muscle properties as observed through EMG: chronic disuse below the lesion level causes muscle atrophy and alterations in the makeup of the fibers and increased passive stiffening \citep{shields2007musculoskeletal, gorgey2007skeletal}. Clinically, the former is manifested by the abnormal EMG-force relations and increased reflex excitability (spasticity) by the reflexes \citep{grosse2002eeg}. Overall, EMG provides additional information to the EEG in the case of the injured CNS by accounting for the peripheral manifestation resulting following the loss of the central drive ranging through the acute denervation observed in severe strokes to the documentation of residual and long-term persistent neuromuscular adaptation in SCI \citep{heald2017characterization, shields2007musculoskeletal}.

\section{EEG/EMG Changes Following PNS Injury}

In contrast to CNS lesions, pure PNS lesions (e.g., peripheral nerve transections or neuropathies) do not usually induce the diffuse slowening of brain rhythms observed on EEG. This occurs because the intact and functioning brain proper shows no direct cortical destruction to release anomalous delta/theta waves \citep{nuwer1997assessment}. Rather, PNS trauma mainly presents in peripheral electrophysiology, in which nerve conduction studies and EMG offer the most sensitive indices of axonal integrity and reinnervation. Peripheral nerve injury still rests and relies on nerve conduction studies (NCS) and EMG as the most important tools for diagnosis. A peripheral nerve that is completely transected yields no evoked response at the distal muscles and exhibits absent sensory nerve action potentials \citep{kimura2025electrodiagnosis, preston2020electromyography}.

On needle EMG, by 1-2 weeks after the injury, denervation changes such as the presence of fibrillation potentials and positive sharp waves appear in muscles deprived of innervation, confirming the loss of motor axons, similar appearing to denervation patterns following CNS lesions with peripheral expression \citep{kamble2019peripheral, dumitru2002electrodiagnostic}.

By months, successful reinnervation, whether by axonal growth or collateral sprouting, is revealed on EMG by the development of polyphasic MUPs by the presence of newly created neuromuscular links and gradual voluntary motor unit recruitment increase by months \citep{kimura2025electrodiagnosis}.

Of interest, the central consequences of a pure peripheral lesion are restricted. However, peripheral reorganization and plasticity may arise through changing sensory inputs, e.g., following limb transection or nerve lesioning, the somatosensory cortices may become remapped with adjacent body territories growing into the deafferented region \citep{li2021cortical}. In the clinical and experimental realms, SSEPs are employed to follow such alterations. It has long been observed as a classic finding that severe peripheral lesions or transections of the whole spinal cord suppress SEP signals below the lesion level by cutting ascending pathways \citep{nordmark2020disinhibition}. In individuals presenting clinically complete spinal cord injuries, the absence of SSEPs conventionally and reliably represents disruption of ascending afferent pathways and contrasts when preserved or early-terminating SEPs predict the presence of functioning sensory fibers and better neurological recovery and possible recovery \citep{kim2014changes, chrysanthakopoulou2025machine}. For partial peripheral nerve lesions, preserved SEPs are typical but demonstrate delayed latencies or diminished amplitude and indicate slow conducting or diminished numbers of functioning fibers \citep{kamble2019peripheral}. Beyond evoked responses, PNS trauma can furthermore induce long-term cortical adaptations, e.g., individuals may depend more upon visual feedback or contralateral limb control, and FMRIs and EEG studies of connectivity detect nuanced sensorimotor network shifts and not gross impairments \citep{li2021cortical, nordmark2020disinhibition}. Importantly, PNS lesions do not induce the notable global EEG slowing or widespread dissociation among the associations usually observed in the context of CNS lesions; rather, the electrophysiological signature of PNS lesions resides in localized signal loss and EMG evidence for denervation and reinnervation and central brain networks remaining intact apart from nuanced reorganization \citep{preston2020electromyography, nuwer1997assessment}. This difference emphasizes the fact that diffuse EEG/BOLD impairments are the signatures of central injuries and whereas peripheral lesions are most adequately defined by local peripheral electrophysiological alterations.

\section{EEG and EMG Biomarkers During CNS and PNS Regeneration}

During CNS repair, such as following stroke or SCI, EEG measures show typical alterations that mirror functional recovery. One reliable finding is spectral balance shift, characterized by decreased pathological slow-wave activity and relative enhancement of the alpha- and beta-band rhythms, while motor function contributes and improves \citep{wu2015connectivity}. Moreover, movement-related EEG dynamics during event-related desynchronization (ERD) in the mu (8-12 Hz) and beta (13-30 Hz) bands across the sensorimotor cortex become greater when patients recover voluntary motion \citep{pichiorri2015brain}. Such EEG alterations mirror intrinsic cortical reorganization: greater ERD/ERS response strength reflects greater reactivation of the lesioned sensorimotor cortex while better interhemispheric beta coherence reflects recovery of more physiological oscillatory coordination among motor areas \citep{wu2015connectivity, bonstrup2018parietofrontal}. Supplementing such EEG measures, EMG offers direct evidence of neuromuscular recovery in stepwise correspondence with central reorganization. By surface EMG measures, the re-emergence and enhancement of motor unit activity in paretic muscles can be tracked. Following stroke, gradual EMG amplitude and complexity increases during movement attempts signify that recruitment and discharge rates of motor units are recovering and reflect the patient's improving capacity for voluntary contraction and finer neuromuscular coordination. Note that by EMG-derived measures such as amplitude and motor unit complexity precise correlations exist across muscle strength gains and functional motor recovery \citep{li2021cortical, al2023electromyography, stefanovic2022rate}. In SCI, EMG measures are delicate enough to discern residual or re-activating muscle activation even when overt movement cannot be observed and may consequently provide evidence for subclinical re-activation of motor pathways below the lesion. In the process of rehabilitation, muscles that appear latent may begin to demonstrate EMG activity by small, early MUPs. This represents the preliminary evidence of reinnervation and corticospinal reconnection prior to the development of noticeable movement \citep{balbinot2022surface, heald2017characterization, shields2007musculoskeletal}.

Joint EEG-EMG analyses provide valuable indications that reveal both central and peripheral aspects of nerve recovery. One widespread indicator is cortico-muscular coherence (CMC), which examines the degree to which brain waves (EEG) and muscle activities (EMG) collaborate. Following brain lesions such as a stroke, CMC, particularly the beta range, is sharply decreased, displaying weak associations among the brain and spinal cord. But during the process of recovery and retraining, the values of CMC gradually increase, indicating that the association among the motor cortex and the muscles improves \citep{krauth2019cortico, mima1999corticomuscular}. Early research on recovery following a stroke revealed that EEG-EMG coherence during the movement at the wrist was low at the beginning and then rose as motor abilities grew and sometimes reached or exceeded values observed in controls \citep{krauth2019cortico, pichiorri2023exploring, mima2001coherence}. Such findings confirm the hypothesis that greater associations among muscles and the brain are a significant indicator of nerve reorganization. When performing motor tasks, synchronized EEG and EMG responses are evident. For instance, a noticeable EEG desynchronization occurring concurrently with an EMG burst during voluntary movement reveals that muscle output and brain activation collaborate effectively in the recovering system. Such synchronized EEG-EMG responses, including movement-related beta-band ERD occurring concurrently with muscle activity revealed by EMG, demonstrate the gradual re-establishing of feedback and feedforward loops among the CNS and the muscles \citep{krauth2019cortico, mima1999corticomuscular, pichiorri2023exploring}.

In the PNS, EMG continues to be a key biomarker of regeneration. Peripheral nerve transections or compressions induce muscle denervation, but growing axons can ultimately reinnervate muscle fibers. Needle EMG offers some of the first objective evidence of the process, and the observation of small, polyphasic MUPs represents a signature of early reinnervation \citep{kamble2019peripheral, kimura2025electrodiagnosis}. These new polyphasic MUPs, usually observed in the first few months following injury, indicate that newly growing axons have achieved new neuromuscular junctions and began to activate muscle fibers even when contractions are too feeble to create discernible movement. As reinnervation progresses, EMG patterns change: MUPs become larger and longer in duration, reflecting the enlargement of motor units through collateral sprouting, and recruitment of motor units grows stronger and more coordinated \citep{sulaiman2013neurobiology, kamble2019peripheral, kimura2025electrodiagnosis}. At the same time, voluntary EMG activity grows stronger and stronger. EMG can be used then as a marker of neuromuscular recovery, appearing early, usually detecting enhancement in activation of motor units and neuromuscular transmission prior to clinical strength appearing by observation \citep{sadek2024decomposition, rodney2024neuromuscular}. Overall, the combination of EMG and EEG allows for a better understanding of the picture of regeneration, EEG shows how the cortical activity reorganizes, and the EMG confirms the peripheral implementation of those recovering motor signals, ranging from the reemergence of the MUPs to the construction of the new muscle contractions.

\section{EEG and EMG Alterations in PNS and CNS Injury and Recovery}

Peripheral nerve injury disrupts the ongoing exchange of information between the brain and the limb and produces radical alterations in cortical processing. EEG research shows that the disruption of afferent and efferent input with peripheral trauma initiates sensorimotor cortical areas' neuroplastic reorganization (e.g., remapping, altered connectivity) \citep{fraiman2016reduced}. This disruption can be revealed as a shift in oscillatory activity in the cortex, e.g., spectral power or stability change or as reduced activation in the cortical regions innervated by the damaged nerve and reflective of the adaptive reorganization by the brain following deafferentation. Meanwhile, the electromyographic (EMG) records provide electromyographic evidence of peripheral denervation: voluntary MUPs disappear soon after trauma onset, and involuntary fibrillation potentials, and positive sharp waves, typically become evident soon after two to three weeks following trauma onset \citep{kamble2019peripheral, han2020pre}. These respective neurophysiological findings collectively outline the central and peripheral interactions soon after nerve trauma.

During the recovery phase when peripheral nerves reinnervate target muscles and refill, both EEG and EMG measures change to reflect functional recovery. More specifically, recovery of afferent sensory feedback and motor drive enables cortical re-activation; formerly-suppressed cortically active areas may demonstrate heightened activation on EEG and reorganization indicative of the recovery of motor control. Moreover, electrophysiological measures by EMG indicate stepwise advances in nerve-muscle transmission; e.g., administration of polyethylene glycol (PEG 3350) in preclinical preparations has been shown to advance recovery by increasing electrophysiological function by shortening compound muscle action potential (CMAP) latency and by boosting amplitude both indicative of functional reinnervation \citep{ni2023electrical, tunc2025harnessing}. Indeed, when a growing nerve connects and develops lasting function, the motor cortex quickly undergoes remodeling to re-acquire effective muscle control by processes of cortical plasticity \citep{assessment2018peripheral, emg2025evaluation}. Meanwhile, EMG demonstrates clear evidence of reinnervation: voluntary motor unit activation re-establishes itself by first appearing as polyphasic, small amplitude MUPs resulting from collateral sprouting, potential values that increasingly mature and develop bigger amplitude and duration as new axons develop synaptic contacts \citep{bateman2025assessment, bateman2023assessment}. Interestingly, the appearance of the polyphasic potentials three to four months following trauma coincides with ongoing functional recovery \citep{bateman2025assessment}. Together, peripheral gains frequently accompany reinforced cortical values on EEG and indicate repair of the PNS-CNS communication circuit. By contrast, CNS lesions such as stroke or SCI provide different but complementary EEG-EMG patterns. At the cortical level, EEG generally shows abnormal sensorimotor rhythms, μ/β ERD alterations during movement attempt or movement imaging, that follow impairment and recovery sub-acute and sub-chronic periods after stroke and appear similarly in SCI subjects \citep{kumari2023predictive, milani2022relation}. At the same time, EMG is indispensable in the evaluation of motor impairment and recovery: surface EMG can reveal residual voluntary muscle output in paretic muscles and even assist movement intent decoding when overt movement is small \citep{meyers2024decoding}. It also defines spasticity and coordination, objective sEMG measures such as RMS/iEMG and the index of co-contraction measure hypertonia and pathological co-activation and display correlations with clinical rating scales (e.g., MAS) in the patient populations for post-stroke or SCI \citep{leszczynska2023unveiling, xie2024mapping}. The patterns of pathological muscle co-activation are quantifiable: EMG indices of co-contraction separate individuals after stroke from controls and can follow the intervention-related alterations and demonstrate the utility for evaluating coordination impairments \citep{bandini2023surface}. The combination of EEG and EMG provides further information. The simultaneous registration demonstrates decreased cortico-muscular coherence (EEG-EMG synchrony) following stroke but increasing cortico-muscular coherence correlating with motor recovery and proving to be a marker of the restored corticospinal coupling \citep{krauth2019cortico, pichiorri2023exploring}.

\begin{table}[htbp]
\centering
\caption{Timeline of Electrophysiological Changes in Neural Injury and Recovery}
\label{tab:timeline}
\resizebox{\textwidth}{!}{%
\begin{tabular}{|p{2.5cm}|p{4.5cm}|p{4.5cm}|p{3.5cm}|}
\hline
\textbf{Time Period} & \textbf{CNS Injury} & \textbf{PNS Injury} & \textbf{Recovery Phase} \\
\hline
\textbf{Acute} \\(0-2 weeks) 
& \textbf{EEG:} Immediate slowing, disrupted networks \newline \textbf{EMG:} Reduced recruitment, possible denervation in severe cases 
& \textbf{EEG:} Minimal changes \newline \textbf{EMG:} Initially normal due to remaining ACh, then progressive loss of responses 
& Assessment of injury severity and prognosis \\
\hline
\textbf{Subacute} \\(2 weeks - 3 months) 
& \textbf{EEG:} Continued slowing, network reorganization begins \newline \textbf{EMG:} Spasticity development, abnormal patterns 
& \textbf{EEG:} SEP changes appear \newline \textbf{EMG:} Fibrillation potentials and positive sharp waves emerge (week 2-3) 
& Early plasticity and potential for intervention \\
\hline
\textbf{Chronic} \\(3+ months) 
& \textbf{EEG:} Gradual normalization with recovery \newline \textbf{EMG:} Improved recruitment, reduced co-contraction 
& \textbf{EEG:} Cortical remapping \newline \textbf{EMG:} Polyphasic MUPs appear, progressive maturation 
& Functional recovery and long-term adaptation \\
\hline
\end{tabular}%
}
\end{table}

Mechanically, a higher level of abnormal co-contraction reflects corticospinal tract dysfunction and hence verifies the status of the corticospinal tract as a pathophysiologically anchored measure in stroke rehabilitation \citep{sheng2022upper}. Likewise, the recovery of time-locked movement-related EEG responses, like movement-related cortical potentials (MRCPs) and $\mu$/$\beta$ ERD, reflect the re-establishment of motor pathways. Longitudinal EEG investigations demonstrate that premovement delta/β oscillations recur in association with behavioral recovery after a stroke, while preceding assessment of ERD may provide predictors of later motor recovery \citep{antonioni2024event}. Additionally, MRCPs during lower-limb tasks in stroke patients efficiently monitor the execution and timing of gait phases and support the value of MRCPs as time-locked measures of restored motor execution and planning \citep{phang2024time}. Together, the combination of EEG-EMG modes (e.g., cortico-muscular coherence extracted by using EEG-EMG concurrently) underscores the central-peripheral relationship at the dynamic level and is employed for the development of neurorehabilitation techniques, such as the deployment of hyrid BCI techniques \citep{pichiorri2023exploring}.

\section{Discussion}

EEG in both CNS and PNS injuries routinely demonstrates utility as a non-invasive, dynamic indicator of network disruption and recovery. For CNS injuries like stroke or SCI, EEG generally presents widespread slowing and desynchronization in affected networks. Following stroke, rest-state recordings routinely display decreased alpha/beta power by accompanying elevations in theta/delta over the lesioned hemisphere, and aggregate qEEG indices (e.g., delta-alpha ratios) relate to clinical condition in acute-subacute periods \citep{zhang2024resting, sood2024quantitative}. In SCI, sensorimotor rhythm alterations and alpha/theta metric shifts are also implicated, confirming the validity of using EEG to monitor impairment and rehabilitation-related alterations \citep{simis2021electroencephalography, albuquerque2023modulation}. EMG complementarily describes the end-stage motor outcomes of central lesions and demonstrates decreased voluntary motor-unit recruitment, abnormally irregular activation timing throughout task phases, excessive agonist-antagonist co-contraction and spastic bursts, in concordance with compromised descending drive and maladaptive reflex control \citep{bitencourt2022needle, bandini2023surface, hohsoh2024comparative}.

In contrast, for PNS injuries (e.g., peripheral nerve transection/entrapment), the cortex is intact structurally and EEG alterations arise mainly through deafferentation and disuse, rather than direct cortical lesion; experimental and human research confirm sensorimotor-cortical hyperexcitability/reorganization motivated by loss, and restitution, of peripheral input \citep{nordmark2020disinhibition, meijs2024spared}.

SEPs allow tracking the somatosensory pathways and generally demonstrate diminished/absent response findings in the acute period following root/nerve compression and improve when function returns or when compression is relieved by the release; SEP amplitude and time-frequency power scale by lesion size and normalize when function returns \citep{sun2023integration, jo2021diagnostic}.

Concurrently, EMG assumes the major physiological endpoint of peripheral integrity. Early after axonal loss, needle EMG demonstrates denervation, fibrillation potentials, and positive sharp waves. In contrast, with reinnervation, new polyphasic, low-voltage motor-unit potentials appear and then increase in size and duration as collateral sprouting and axonal growth restore motor units \citep{krarup2016remodeling, debenham2025evaluating, fidanci2023needle}.

However, intense PNS deafferentation can bring about cortical rhythm disorders similar to those following CNS lesions; in chronic SCI populations, the resting EEG demonstrates shifts to lower frequencies and changed beta activity, characteristics typical for thalamocortical dysrhythmia \citep{simis2021electroencephalography, boord2008electroencephalographic}. Overall, CNS lesions usually result in gross EEG change, and PNS injuries mainly activate focal cortical reorganization, indicative of the plastic response by the brain to the peripheral input change.

Most importantly, EEG biomarkers are related to neuroplastic transformation and functional restoration. Larger delta-alpha and DTABR ratio and lower alpha/beta power index are connected with greater impairment and tissue hypoperfusion in subacute ischemic stroke, and longitudinal studies demonstrate normalization of the measures with recovery progress \citep{ajcevic2021early, sood2024quantitative}. Task-related EEG also shows the re-establishment and enhancement of movement-related ERD/ERS in lockstep with motor recovery. Similarly, EMG signatures change with recovery: voluntary amplitudes rise as more graded recruitment improves; antagonist co-contraction diminishes as the subject undergoes rehabilitation; and new, small-amplitude polyphasic motor-unit potentials evolve to larger, long-duration, more stable units as collateral sprouting and new axonal growth continue \citep{bandini2023surface, negro2020impaired}.

In SCI recovery, parietal/sensorimotor alpha/theta ratio increases are related to better gait/balance function; qEEG studies show the alpha/theta ratio reflects the progression of functional recovery during the treatment process, and beta power in relation to tasks also signifies gait recovery in sub-acute SCI \citep{simis2021electroencephalography, simis2020beta}. At the same time, beta-band cortico-muscular coherence (CMC), typically decreased shortly after stroke, increases during the process of the recovery period and associates with the recovery of effective corticospinal drive and motor control \citep{krauth2019cortico, delcamp2023corticomuscular, parmar2024beta}. Correspondingly, in stroke, EEG indices of functional connectivity show the strengthening of intra-hemispheric coupling and reduction of maladaptive inter-hemispheric dominance in damaged motor networks during the recovery process, comparable to regained function. Coherence and task-EMG patterns by EMG typically mirror EEG network changes and provide peripheral evidence that increased cortical coordination produces improved muscle activation; aberrant EEG-EMG coupling (reduced beta-band cortico-muscular coherence) augments during the process of motor recovery and increased coordination in sub-acute-stroke groups \citep{zhang2024resting, chen2023assessment}.

In sync, EEG marker refinement, increased alpha power, recovery of inter-areal connectivity, and greater movement-related ERD upon attempted action, reflect functional recovery and upper-limb function \citep{zhang2024resting, xu2024event}. Coincident EMG measures demonstrate greater, more selective recruitment and less co-contraction with practice, whereas increasing cortico-muscular coherence measures the re-establishing effective corticospinal drive \citep{bandini2023surface, chen2024change}. In conjunction, the EEG-EMG signatures match network-level reorganization assessed by measures of connectivity and transcranial magnetic stimulation (TMS)-EEG and concordant re-establishment of functional circuits \citep{bai2023intracortical, pirovano2022resting}. EEG offers the following several distinctive merits in the monitoring of neural regeneration: it measures directly the neural activity at the millisecond scale, is portable/inexpensive (including bedside and prehospital), and may be repeated frequently to derive moment-to-moment plasticity; besides, quantitative EEG indices (e.g., DAR, DTABR, alpha/beta power) offered prognostic information on disability and outcomes following stroke \citep{sood2024quantitative, shen2024predicting, wilkinson2020application}.

EMG also offers a bedside, repeatable measure of functional output, converging to confirm that central gains are translating to muscle activation, picking up early reinnervation following peripheral repair (emergence of fibrillation potentials/positive sharp waves by $\sim$2-3 weeks), and measuring abnormal synergies or spasticity in the CNS injury groups \citep{kamble2019peripheral, leszczynska2023unveiling}. Clinically, a beneficial resting EEG pattern (e.g., reduced DAR or recovered alpha/beta) may recommend progression to higher intensity training, while abnormally ongoing rhythms necessitate change of approach; concurrently, the onset of voluntary EMG activity or decreased recruitment threshold provides support for progression to greater difficulty tasks and selective muscle re-education by target muscles \citep{wang2021quantitative}. Importantly, both EEG and EMG alterations often antedate explicit behavioral improvements, providing an early insight as to whether circuits are going the right way functionally \citep{kamble2019peripheral}.

Clinical applications of EEG biomarkers in real life are gaining pace. Neurorehabilitation uses qEEG measures (e.g., delta-alpha ratio, DTABR, alpha/beta power) as surrogate end-points to monitor brain alterations related to therapies and guide prognosis, retrospective and prospective studies demonstrate that elevated slow/fast ratios and reduced alpha/beta in the acute period after stroke predict poor outcomes, whereas normalization accompanies recovery \citep{shen2024predicting, wang2021quantitative, benghanem2024prognostic, zhang2025retrospective}. Most potentially transformative is the combination of EEG and EMG in brain-computer interfaces (BCIs) and other closed-loop systems. As EEG non-invasively represents motor intent, BCI inputs can control functional electrical stimulation (FES) or exoskeletons, and randomized and controlled trials/meta-analyses document motor improvements in stroke and encouraging findings in SCI \citep{biswas2024single, he2025multimodal}. Dual EMG verifies delivery and provides biofeedback/device control; EMG-activated stimulation protocols (e.g., EMG-neuromuscular electrical stimulation (NMES)/contralaterally controlled neuromuscular electrical stimulation (CCNMES)) enhance upper-limb function and cortical activation compared with standard stimulation in systematic reviews and controlled studies \citep{yang2023effect, shen2022effectiveness}.

Following PNS injury/nerve transfer, EMG-activated stimulation facilitates muscle preservation and retraining while reinnervation advances, and myoelectric prostheses (after often preceding targeted muscle reinnervation) allow functional control during the recovery period \citep{lin2025distal, costello2023clinical}. Ancillary sensory pathway tracking by means of SEPs records loss and recovery of afferent conduction as function returns \citep{fuseya2025somatosensory}. Such closed-loop EEG-EMG systems functioning altogether serve as powerful therapies' adjuncts, transcribing central intent to peripheral effectors and providing feedback proprioceptive input to support use-dependent plasticity \citep{ferrero2023brain}. Although powerful, EEG biomarkers are not without limitations: the spatial resolution is weak and deep/focal generators are still difficult to localize; estimates are heavily reliant on preprocessing and feature selection; inter-subject difference is non-trivial; and signals are very vulnerable to motion, muscle, and ocular artifacts \citep{kuroda2024test, gomez2025evaluation}. EMG also has limitations: since it is an end-organ signal, it cannot alone distinguish central drive failure and peripheral conduction block, and amplitude is not a direct measure of neural drive (e.g., amplitude-cancellation effects). It is vulnerable to crosstalk and electrode positioning, and confounded by fatigue; single-muscle records omit compensatory patterns, although thoughtful protocols and high-density arrays reduce some difficulties \citep{bedoy2025improving, wimalasena2022estimating}. Therefore, electrophysiological biomarkers should be an adjunct, not substitute, for the clinical examination and imaging when informing diagnosis and rehabilitation decisions \citep{hirsch2021american}. please see Table \ref{tab:advantages_limitations}

To compensate for such limitations, multimodal systems couple EEG with complementary techniques. EEG-fMRI brings millisecond dynamics together with spatially mapped measures (e.g., wholesale activation or deactivation) to allow for targetable, biomarker-guided rehab and neurofeedback in stroke populations. EEG + diffusion MRI (DTI) references oscillatory modifications relative to corticospinal-tract integrity (e.g., FA), the latter tracking motor recovery, referencing network physiology to structural substrate \citep{butet2025eeg}. EEG-EMG coupling through cortico-muscular coherence actually measures descending pathway function and restoration through rehabilitation in real time \citep{parmar2024beta}. Similarly, hybrid EEG-fNIRS systems enhance motor intent and recovery tracking through the addition of hemodynamic context to the EEG timing/frequency content \citep{li2023early}. Lastly, TMS-EEG/EMG protocols interrogate circuit excitability and plasticity longitudinally, TMS-evoked EEG potentials measure cortical reactivity/connectivity while MEPs report corticospinal drive during recovery \citep{bai2023intracortical}. In general, multimodal monitoring links single-modality gaps: while EEG lacks good spatial resolution, imaging localizes; while EEG is indefinite, EMG or hemodynamics verify; and while EMG displays output without central signatures, EEG unravels intent and network state. For the future, multiple priorities may allow for the fullest exploitation of the utility of EEG (and EMG). First, harmonization of electrophysiological biomarkers, from spectral proportions (e.g., Delta/Alpha Ratio (DAR) and Delta + Theta / Alpha + Beta Ratio (DTABR)) and indices of connectivity to beta-band cortico-muscular coherence and motor-unit measures, would elevate comparability and clinical translation; recent studies in stroke demonstrate that qEEG ratio measures mirror prognosis, highlighting the utility in agreed-on definitions and pipelines \citep{sood2024quantitative, wang2021quantitative}. Second, enhanced spatial mapping through high-density EEG and advanced-source models, and by high-density EMG (HD-EMG) for motor-unit decomposition, can strengthen links between physiology and anatomy and measure reinnervation at the motor-unit level \citep{grison2025unlocking}. Third, longitudinal multimodal cohorts that match the trajectory of EEG/EMG to imaging (e.g., DTI of the corticospinal tract), nerve-conduction/SEP measures, and outcomes must be undertaken to elucidate how electrophysiological recovery reflects tissue regeneration at the whole-organism level \citep{schulz2021corticospinal}. Fourth, BCI developments, such as adaptive EEG-EMG control and closed-loop stimulation, should focus on maladaptive patterns and scale to home use; nascent work using multiple EEG systems and portable HD-EMG control demonstrates the promise for home assistance and therapy \citep{yang2025high, qiu2022improved}. Finally, broadening beyond the motor endpoints to sensory systems (e.g., using SEPs to mark peripheral recovery) and the pain/cognition realms will increase breadth, of which SCI may be the greatest beneficiary since oscillatory EEG features are related to the neuropathic-pain phenotype and may be tracked longitudinally \citep{nawaz2024electroencephalography, chiu2024lower}. The advantages and limitations of EEG and EMG as neural regeneration biomarkers are detailed in Table~\ref{tab:advantages_limitations}.

\begin{table}[htbp]
\centering
\caption{Advantages and Limitations of EEG and EMG as Neural Regeneration Biomarkers}
\label{tab:advantages_limitations}
\resizebox{\textwidth}{!}{%
\begin{tabular}{|p{2.5cm}|p{6cm}|p{6cm}|}
\hline
\textbf{Modality} & \textbf{Advantages} & \textbf{Limitations} \\
\hline
\textbf{EEG} 
& \begin{itemize}
\item Non-invasive and portable
\item High temporal resolution (milliseconds)
\item Real-time monitoring capability
\item Cost-effective and repeatable
\item Bedside application
\item Direct neural activity measurement
\end{itemize}
& \begin{itemize}
\item Poor spatial resolution
\item Susceptible to artifacts (muscle, eye, motion)
\item Difficulty localizing deep sources
\item Inter-subject variability
\item Requires expertise for interpretation
\end{itemize} \\
\hline
\textbf{EMG} 
& \begin{itemize}
\item Direct measure of motor output
\item Early detection of reinnervation
\item Objective assessment of function
\item Guidance for rehabilitation
\item Quantification of motor unit properties
\end{itemize}
& \begin{itemize}
\item End-organ signal only
\item Cannot distinguish central vs. peripheral causes
\item Crosstalk between muscles
\item Position-dependent recordings
\item Limited to accessible muscles
\end{itemize} \\
\hline
\textbf{Combined EEG-EMG} 
& \begin{itemize}
\item Comprehensive central-peripheral view
\item Cortico-muscular coupling assessment
\item Enhanced diagnostic accuracy
\item Better understanding of recovery mechanisms
\end{itemize}
& \begin{itemize}
\item Increased complexity
\item Higher technical requirements
\item More susceptible to artifacts
\item Requires specialized analysis
\end{itemize} \\
\hline
\end{tabular}%
}
\end{table}

By moving forward with standardization, spatial mapping, and longitudinal multimodal design, future research will enhance the validity and interpretability of electrophysiological biomarkers and integrate them more robustly in research and clinical care. As part of this system, EEG will measure cortical reorganization and network connection and EMG will confirm peripheral execution and reinnervation, ultimately delivering a unified readout of CNS and PNS regeneration.

\section{Conclusion}

In short, EEG and EMG collectively offer a complementary view into the process of neural regeneration, both the real-time electrical signature of the brain and the peripheral manifestation of motor commands as the PNS and CNS respond and readjust to trauma. For both CNS and PNS situations, such biomarkers represent key processes of recovery: EEG identifies the re-establishment of balanced oscillations, coupling, and cortical re-mapping by the brain while EMG identifies the recovery of voluntary muscle activation, motor-unit recruitment refinement, and cortico-muscular coherence restoration. Together, they tie the activity at the level of the neuron to explicit functional outcomes, reflecting the reorganization of circuits that constitutes the recovering abilities of the patient.

As non-invasive, economical, and real-time monitoring agents, EEG and EMG provide clinicians and investigators with a valuable tool set to observe neural restoration in progress and to offer real-time guidance for rehabilitation. Together, the two can monitor the path of recovery, provide prognosis guidance, and individualize therapy, whether by adapting a regimen in response to changing brain rhythms, by modulating exercises in response to recruitment patterns by muscle, or by using closed-loop BCI and FES systems to actively recruit latent pathways. This multi-modality method applies generally across central (stroke, TBI, SCI) and peripheral (nerve lesions, transections) lesions and thus applies generally wherever restoration of the neurons or the circuits they compose is desired.

In the future, as technology advances and multimodal techniques become more common, EEG-EMG integration will become an increasingly powerful biomarker system, yet one that not only observes regeneration but actually promotes it. Embracing both modalities in science and clinical practice will continue to expand our ability to both improve recovery and take advantage of the brain's and muscles' own messages to enlighten and hasten the process of repair at the neural level.

\bibliographystyle{unsrt}
\bibliography{ref}

\end{document}